# Non-equilibrium Magnetic Properties of Single Crystalline $La_{0.7}Ca_{0.3}CoO_3$

Asish K. Kundu[a,b], P. Nordblad[a] and C. N. R. Rao[b*]

[a]Department of Engineering Sciences, Uppsala University, 751 21 Uppsala, Sweden

[b]Chemistry and Physics of Materials Unit, Jawaharlal Nehru Centre for Advanced Scientific Research, Jakkur P.O., Bangalore-560064, India

**Abstract**

Magnetic and electric properties of a single crystal of $La_{0.7}Ca_{0.3}CoO_3$ have been experimentally studied. The system attains a ferromagnetic spontaneous moment below 170 K and exhibits a re-entrant spin-glass phase below 100 K. In the ordered and the re-entrant phases, the low field magnetic properties are strongly direction dependent, showing considerably higher magnetization values perpendicular than parallel to the c-axis. Magnetic relaxation experiments show that both the ferromagnetic and the re-entrant spin-glass phases are non-equilibrium states, where the system exhibits magnetic aging characteristic of spin-glasses and disordered and frustrated ferromagnets.

*for correspondence: cnrrao@jncasr.ac.in

## I. INTRODUCTION

Rare-earth manganates of the general formula $Ln_{1-x}A_xMnO_3$ (Ln = rare earth, A = alkaline earth), possessing the perovskite structure, exhibit interesting properties such as colossal magnetoresistance, charge ordering and electronic phase separation.[1-5] These properties of the manganates are strongly influenced by the average radius of the A-site cations, $<r_A>$. Perovskite cobaltates of the type $Ln_{1-x}A_xCoO_3$ (Ln = rare earth, A = alkaline earth) are somewhat similar to the manganates.[6-8] Accordingly, many of the cobaltates show ferromagnetism and metallicity depending on the composition and the size of the A-site cations. The cobaltates also show some unusual features in their magnetic properties. Thus, Itoh et al[9] report a spin-glass behaviour in $La_{1-x}Sr_xCoO_3$ when $0.0 \leq x \leq 0.18$ and a cluster-glass behaviour when $0.18 \leq x \leq 0.5$. Ganguly et al[10] reported long-range freezing of superparamagnetic clusters when $x < 0.3$. For $x = 0.5$, a cluster-glass magnetic behaviour has been reported by Kumar et al[11] who ascribe this property to magnetocrystalline anisotropy. Wu et al[12] suggest that $La_{1-x}Sr_xCoO_3$ is best described as one dominated by glassy ferromagnetism and magnetic phase separation.[13] Similarly, Burley et al[14] reported long-range ferromagnetism and glassiness in $La_{1-x}Ca_xCoO_3$, as well as a structural phase transition for $x > 0.1$. A recent investigation of the magnetic and electric properties of polycrystalline samples of $La_{0.7-x}Ln_xCa_{0.3}CoO_3$ (Ln = Pr, Nd, Gd or Dy), reinforces the phase separation scenario wherein large carrier-rich ferromagnetic clusters and carrier-poor smaller clusters coexist.[15] The parent compound $La_{0.7}Ca_{0.3}CoO_3$, itself shows the divergence between zero-field-cooled (ZFC) and field-cooled (FC) magnetization data and the $M_{FC}(T)$ curve does not show saturation down to low temperatures. In order to fully understand the unusual behaviour of $La_{0.7}Ca_{0.3}CoO_3$, we have carried out detailed magnetic measurements on single crystal samples at very low fields and also along different directions. More importantly, we have studied the magnetic relaxation to throw light on the nature of phase separation. The study has revealed that the ferromagnetic phase below 170 K is a non-equilibrium phase similar to that observed in prototype re-entrant ferromagnets while below 100 K a low-temperature re-entrant spin-glass phase emerges.

## II. EXPERIMENTAL PROCEDURE

A floating-zone furnace fitted with two ellipsoid halogen lamps with radiation heating was used to grow the single crystal of $La_{0.7}Ca_{0.3}CoO_3$. Polycrystalline rods (feed and seed) were prepared by conventional solid-state reaction method. Stoichiometric mixtures of the starting materials $La_2O_3$, $CaCO_3$ and $Co_3O_4$ were weighed in the desired proportions and milled for a few hours. After the mixed powder was dried, it was calcined in air at 1223 K and after few intermediate grinding the powder was finally sintered at 1473 K for 24 h in air. The sample was then reground and mono-phasic polycrystalline powder was hydrostatically pressed and sintered at 1473 K for 24h in air to obtain feed and seed rods with a diameter of 4 mm and a length of 100 mm. A single crystal was then grown under an oxygen flow of 3-lit/min at a growth rate of 7 mm/h. A part of the crystal was cut off and ground to a fine powder on which an X-ray diffractogram was measured using a Seifert 3000 TT diffractometer, employing Cu-K$\alpha$ radiation. A single-phase diffraction pattern was obtained yielding a rhombohedral[16] structure with R$\bar{3}$C space group and unit cell parameters are $a_r$ = 5.401(4) Å and $\alpha_r$ = 60.1°. The composition has been checked using energy dispersive x-ray analysis (EDX) in a LEICA S440I scanning electron microscope fitted with a Si-Li detector and it confirms the composition within experimental errors. Other pieces of the single crystal rod were cut off for magnetization and resistivity measurements.

A Quantum Design MPMSXL SQUID magnetometer and a non-commercial low field SQUID magnetometer system[17] were used to investigate the magnetic properties of the single crystal. The temperature dependence of the zero-field-cooled (ZFC), field-cooled (FC) and thermoremanent magnetization (TRM) was measured under different applied magnetic fields. Hysteresis loops were recorded at different temperatures in the low-temperature phases of the system. The dynamics of the magnetic response was studied by ac-susceptibility measurements at different frequencies and measurements of the relaxation of the low field ZFC magnetization.

In the measurements of the temperature dependence of the ZFC magnetization, the sample was cooled from 320 K to 5 K in zero-field, the field applied at 5 K and the magnetization recorded on re-heating the sample. In the FC and TRM measurements, the sample was cooled to 5 K in an applied field and the magnetization recorded on re-heating the sample, keeping the field applied (FC procedure) or after cutting the field off (TRM procedure). In the relaxation experiments, the sample was cooled in zero-field from a reference temperature of 170 K to a measuring temperature, $T_m$, and kept there during a wait time, $t_w$. After the wait time, a small probing field was applied and the magnetization recorded as a function of time elapsed, after application of the field. Electrical resistivity ($\rho$) measurements were carried out with a standard four-probe method in the 20-300 K range with silver epoxy as electrode.

## III. RESULTS AND DISCUSSION

The temperature dependence of the ZFC and FC magnetization measured parallel to the c-axis of the sample at different applied fields (10, 20, 100 and 1000 Oe) is shown in Fig. 1. A sharp increase of the magnetization occurs in the low-field FC data around 170 K indicating that the system orders and attains a spontaneous magnetization. A characteristic strong irreversibility between the low field ZFC and FC magnetization curves appears just below the transition temperature. The ZFC data show a small peak close to 170 K and a second anomaly around 100 K as a relatively sharp maximum, accompanied by a change in the slope of the FC magnetization curve, indicating a transition in the ordered spin structure. Fig. 2 shows the ZFC, FC and TRM magnetization curves measured in a weak field of 1 Oe. The transition at 100 K appears as a sharp dip in the TRM and a sharp maximum in the ZFC magnetization.

Some of the details of the temperature dependence of $M_{ZFC}(T)$ in Fig. 1 are noteworthy. With increasing field (H = 100 Oe), both the peaks in the ZFC data shift towards lower temperatures. Furthermore, on increasing the field up to 1000 Oe, the high temperature $M_{ZFC}(T)$ peak disappears, showing only a shoulder in this temperature region; the 100 K peak becomes broader at higher fields. Another interesting feature is that the irreversibility temperature, $T_{irr}$, (where FC and ZFC curves diverge) decreases with the increasing field, $T_C$ being equal to $T_{irr}$ at sufficiently low fields (10 and 20 Oe), however for higher fields $T_{irr} < T_C$. When the applied field is increased, the $M_{FC}(T)$ and $M_{ZFC}(T)$ curves merge down to ~ 100 K ($T_f$), a behaviour akin to that of ordinary spin-glasses.[18]

In Fig. 3, the magnetization data at 1000 Oe are plotted in the temperature range of 150-320 K along with the high-temperature inverse magnetization data (see inset). The sample shows Curie-Weiss behaviour in the 200-300 K range and a fit to the Curie-Weiss law yields a ferromagnetic Weiss temperature, $\theta_p$, of 150 K and a $p_{eff}$ of ~ 1.6 $\mu_B$/Co-ion, implying S ~ 0.6 $\mu_B$. The derived Weiss-temperature is lower than the ferromagnetic transition temperature (170 K), defined from the inflection point of the inverse magnetization vs. temperature curves. The anomaly occurring around 100 K can be considered to represent a re-entrant spin-glass transition, similar to that found in $Y_{0.7}Ca_{0.3}MnO_3$ by Mathieu et al.[19]

In Fig. 4, we show M-H loops measured parallel to the c-axis at different temperatures. A large hysteresis loop develops below the transition temperature at 170 K and attains a remanence value of ~ 0.6 $\mu_B$/Co-ion and a coercive force of ~ 0.6 Tesla (T) at the lowest temperature. The material becomes harder with decreasing temperature, with the coercive field increasing monotonically with decreasing temperature from 0.02 T at 160 K (near $T_C$) to 0.6 T at 10 K. At temperatures higher than $T_C$ (T = 250 K), the M-H behaviour is linear corresponding to a paramagnetic state.

Figure 5 presents the $M_{FC}(T)$, $M_{ZFC}(T)$ and M-H behaviour for two different orientations of the crystal (H || c-axis and H ⊥ c-axis). The magnitudes of $M_{FC}(T)$ and $M_{ZFC}(T)$ are both higher in the perpendicular direction. Whereas the shape of $M_{ZFC}(T)$ curve remains similar, the shape of $M_{FC}(T)$ changes considerably depending on the direction. In spite of the large c-axis dependence of the initial low-field behaviour, the high field part of the M-H curves is almost independent of field direction as can be seen from Fig. 5 (b). The coercive field and remanent magnetization are almost the same for both the orientations; the magnetic moment at 5 T is the same (~ 0.9 $\mu_B$/Co-ion).

Figure 6 shows the in-phase $\chi'(T)$ and out-of-phase $\chi''(T)$ components of the ac-susceptibility below Curie temperature ($T_c$). A rather similar behaviour has also been reported in some manganate systems by Nam et al.[20] The in-phase component shows, in accordance with the low field ZFC magnetization, two distinct peaks: a frequency-independent high temperature peak at about 170 K that indicates a ferromagnetic ordering and a lower temperature at 100 K peak that is sharper and frequency-dependent. The frequency-dependence of the ac-susceptibility near and below the 100 K anomaly is stronger than at higher temperatures.

In order to explore whether the system is in a thermodynamic equilibrium state or in a non-equilibrium spin-glass like state, the relaxation of the ZFC magnetization was measured at 40, 85 and 110 K after different wait times. The long-time relaxations of the magnetization and aging phenomena well known in spin-glasses[21] are commonly found in many other random magnetic systems. In figures 7, 8 and 9 we present the results of $M_{ZFC}(T,t_w,t)$ and $S(t)$ for $T_m$ = 40, 85 and 110 K, respectively. (It should be mentioned that all the magnetization relaxation curves are measured relative to the first measured data point at $t \approx 0.3$ s, which is set close to zero by a reset of the SQUID electronics). The relaxation rate $S(t)$ is defined by, $S(t) = 1/H\ [dM_{ZFC}(T,t_w,t)/d\log(t)]$, and emphasises the aging features of the relaxation curves. In all the ZFC-relaxation measurements the applied field was 1 Oe and the wait times chosen were $t_w$ = 100, 1000 and 10000 s. The $M_{ZFC}(T,t_w,t)$ measurements show that the sample exhibits logarithmically slow dynamics and an aging effect at all temperatures below $T_c$. The aging phenomenon is revealed from the difference between the $M_{ZFC}(T,t_w,t)$ curves with different $t_w$. We notice a striking aging behaviour, similar to that of spin-glasses at all three temperatures as revealed by an inflection point in the magnetization vs log (t) curves and a corresponding maximum in the relaxation rate curves at an observation time close to the wait time. The relaxation at 40 K is almost two orders of magnitude weaker than the relaxation at the two higher temperatures. In spite of this, the aging character dominates the measured relaxation at all three temperatures. The relaxation of the magnetization measured within our experimental time window (0.3 – 10 000 s) corresponds to a fraction of the total relaxation, and the equilibrium magnetization is far from being reached within this time window. The aging-dominated relaxation observed here is strikingly similar to the behaviour of conventional spin-glasses. In the latter situation, aging is interpreted within the droplet model (or domain growth) model for spin-glasses[22] to reflect the growth of equilibrium spin-glass domains, with the maximum in the relaxation rate being associated with a cross over between quasi-equilibrium (from processes within ordered spin-glass domains) and non-equilibrium dynamics (processes governed by effects at domain walls).

A key property to understand and model the dynamics of spin-glasses is the occurrence of memory. The memory phenomenon is observed in zero-field-cooled magnetization vs. temperature experiments as follows.[23] First a reference experiment is made, according to the ZFC protocol described earlier. A memory curve is then recorded, with the additional feature that the cooling of the sample is halted at a stop temperature for some hours during which the sample ages. This slows down the dynamics of the sample in a region around the stop temperature, which in turn causes a dip in the $M_{ZFC}(T)$ curve. To clearly illustrate memory, it is convenient to plot the difference between the reference and the memory curve. A characteristic of the spin-glass phase (ordinary or re-entrant) is the memory behaviour, whereas a disordered and frustrated ferromagnetic phase would show little or no memory effect. We carried out three experiments, one with a halt below 100 K (85 K), the second with the halt above 100 K (110 K) and the third with halts at both 85 K and 110 K. In Fig. 10 (a), we show the reference curve and the two memory curves. Fig. 10 (b) shows the difference plots. At 85 K, a clear memory dip is observed indicating that the system is in a spin-glass type phase. At 110 K, on the other hand, a memory dip can barely be discerned, indicating that the system is confined in a disordered ferromagnetic phase.

The temperature dependence of electrical resistivity ($\rho$) of the single crystal is shown in Fig. 11. With decreasing temperature there is not much variation in the magnitude of resistivity, but the temperature coefficient of resistivity ($d\rho/dT$) changes rapidly from room temperature to low temperatures. There is a change in slope around 170 K, above which $d\rho/dT$ becomes negative. Below this temperature, $d\rho/dT$ is positive indicating metallicity below $T_c$. Further decreasing the temperature, changes the slope from a positive to a negative value around 80 K, corresponding to an insulating behaviour. In the 20 - 80 K range, the temperature variation of resistivity appears to conform to the variable range hopping regime[24] in accord with the results on polycrystalline samples reported earlier.[15]

## IV. CONCLUSIONS

The main result of the current investigation on single crystals of $La_{0.7}Ca_{0.3}CoO_3$ is that the system enters a non-equilibrium magnetic phase at temperatures below an apparent ferromagnetic transition at 170 K. The non-equilibrium ferromagnetic phase experiences an additional transition into a non-equilibrium re-entrant spin-glass phase at 100 K. The system appears to be phase separated into large carrier-rich ferromagnetic clusters (involves Co in the intermediate-spin (IS) and high-spin (HS) states) and carrier-poor antiferromagnetic or non-ferromagnetic matrix. The M (H) measurements also support this model. The hysteresis curve does not saturate even at higher field 5 T and lowest temperature (10 K). The absence of saturation even at higher fields is a characteristic feature of spin-glass system.[18] Considering the above magnetic behaviour in the low temperature region; we propose that the system goes to re-entrant spin-glass state below 100 K. It is important in this connection to note the unusually large anomalous Hall effect that has been observed at temperatures below $T_C$ in $La_{1-x}Ca_xCoO_3$.[25] The magnetic moment of 1.6 $\mu_B$/Co-ion, found by us is related to the percentage of $Co^{4+}$ in the system. While the $Co^{3+}$ / $Co^{4+}$ ratio is constant through out, the spin-state equilibrium varies with temperature. At low temperatures, both the cobalt ions would be in the low-spin (LS) state, but at higher temperatures the cobalt ions are mostly in the IS and HS states.[26] The value of $p_{eff}$ obtained by us corresponds to a situation where both the $Co^{3+}$ and $Co^{4+}$ ions are in IS and or HS states. The positive $\theta_p$ value can be interpreted in terms of the short-range ferromagnetic interaction between $Co^{3+}$ and $Co^{4+}$ ions, which dominates over the super-exchange interactions between $Co^{4+}$ - $Co^{4+}$ and $Co^{3+}$ - $Co^{3+}$ ions.

## Acknowledgements

Financial support for this work from the Swedish agencies SIDA/SAREC and VR through the Asian – Swedish research links programme is acknowledged. The authors thank BRNS (DAE), India for support of this research. AKK wants to thank University Grants Commission, India for a fellowship award and Victor and Michael for their help during SQUID measurements.

**Figure captions**

**Fig. 1.** Temperature dependence of the ZFC and FC magnetization, M, of $La_{0.7}Ca_{0.3}CoO_3$ at different applied fields (a) H = 10, 20 Oe (b) H = 100 Oe and (c) H = 1000 Oe. ZFC data in open symbol and FC data in solid symbol.

**Fig. 2.** Temperature dependence of TRM, FC and ZFC magnetization, M, recorded for $La_{0.7}Ca_{0.3}CoO_3$ at H = 1 Oe.

**Fig. 3.** Temperature dependence of the ZFC and FC magnetization, M, of $La_{0.7}Ca_{0.3}CoO_3$ (H = 1000 Oe). The inset shows the temperature dependence of inverse magnetization, $M^{-1}$, (H = 1000 Oe).

**Fig. 4.** Typical hysteresis curves for $La_{0.7}Ca_{0.3}CoO_3$ at different temperatures.

**Fig. 5.** Variation of (a) the ZFC and FC magnetization, M, (H = 20 Oe) and (b) magnetic hysteresis curves of $La_{0.7}Ca_{0.3}CoO_3$ for different field directions.

**Fig. 6.** The temperature dependence of the (a) In-phase and (b) Out-of-phase ac-susceptibility of $La_{0.7}Ca_{0.3}CoO_3$ at different frequencies.

**Fig. 7.** ZFC-relaxation measurements on $La_{0.7}Ca_{0.3}CoO_3$ at $T_m$ = 40 K for different waiting times, $t_w$ = 100, 1000 and 10000s (H = 1 Oe).

**Fig. 8.** ZFC-relaxation measurements on $La_{0.7}Ca_{0.3}CoO_3$ at $T_m$ = 85 K for different waiting times, $t_w$ = 100, 1000 and 10000s (H = 1 Oe).

**Fig. 9.** ZFC-relaxation measurements on $La_{0.7}Ca_{0.3}CoO_3$ at $T_m$ = 110 K for different waiting times, $t_w$ = 100, 1000 and 10000s (H = 1 Oe).

**Fig. 10.** ZFC magnetization memory experiment on $La_{0.7}Ca_{0.3}CoO_3$; (a) the temperature dependence of ZFC magnetization, M, (reference curve) and on imprinting memories of two temperature stops (at 85 and 110 K) during cooling each for 3 hours and (b) the difference (M-M$_{ref}$) plot of the respective curves.

**Fig. 11.** Temperature dependence of electrical resistivity, ρ, of La$_{0.7}$Ca$_{0.3}$CoO$_3$.

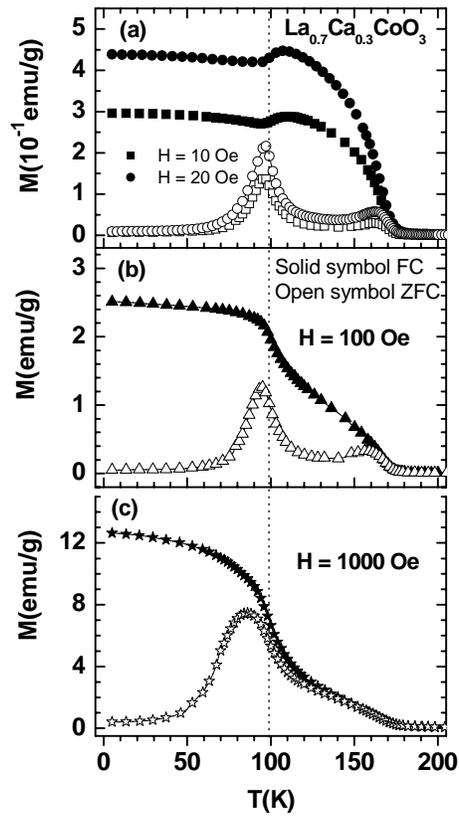

Figure 1

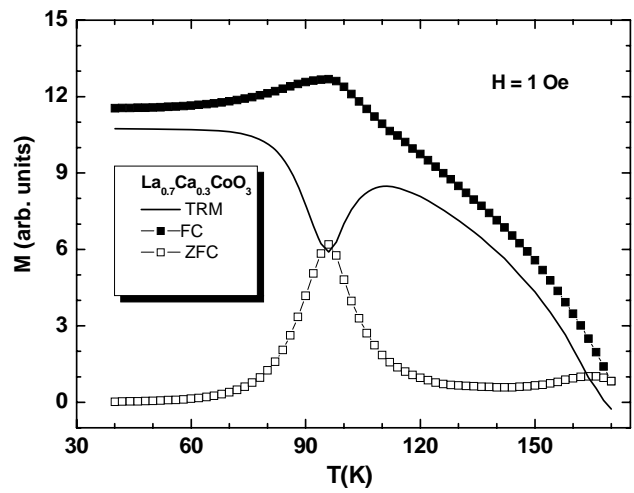

Figure 2

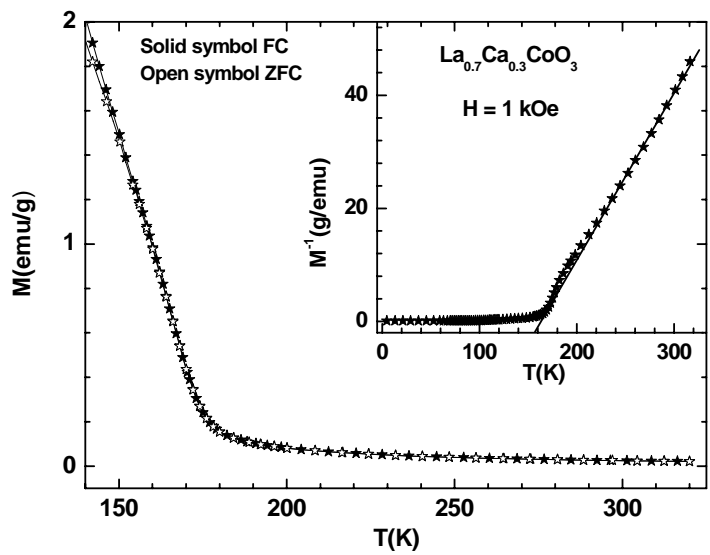

Figure 3.

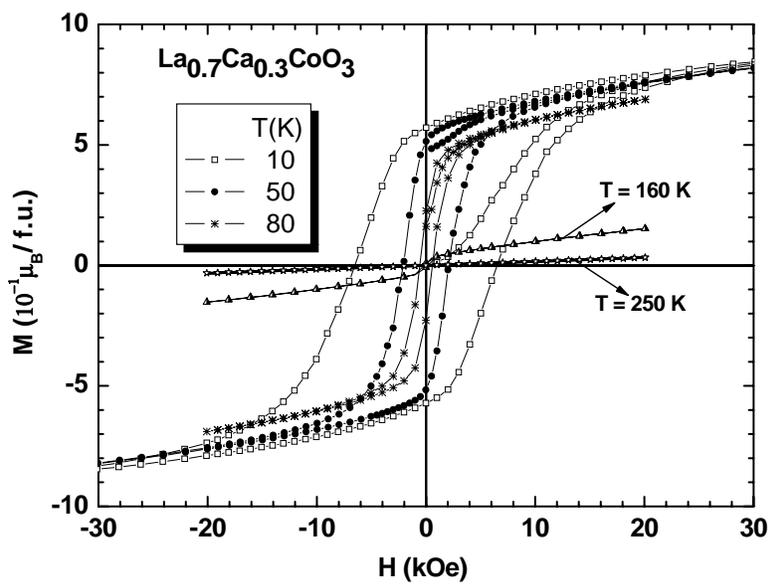

Figure 4.

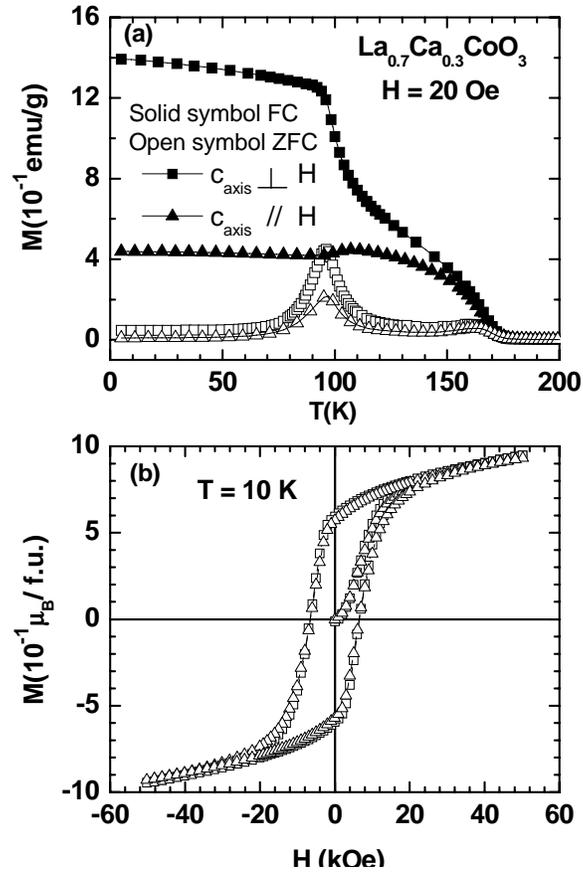

Figure 5.

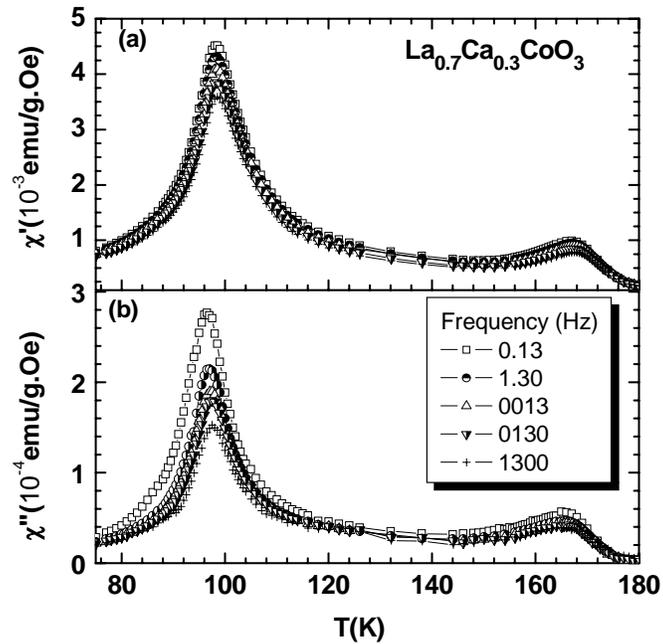

Figure 6.

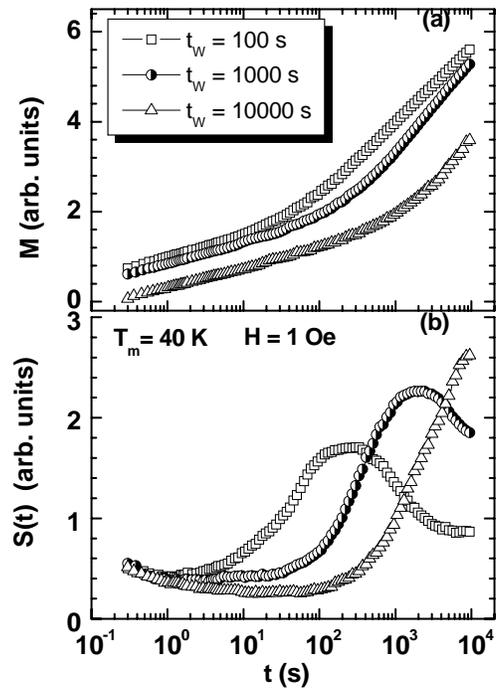

Figure 7.

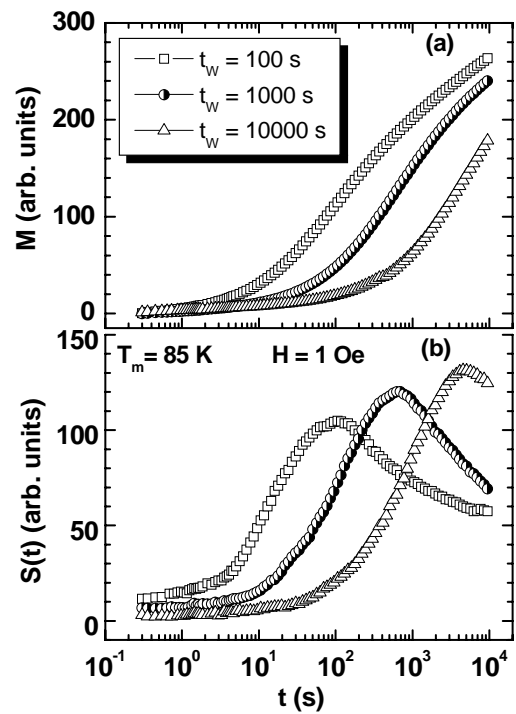

Figure 8

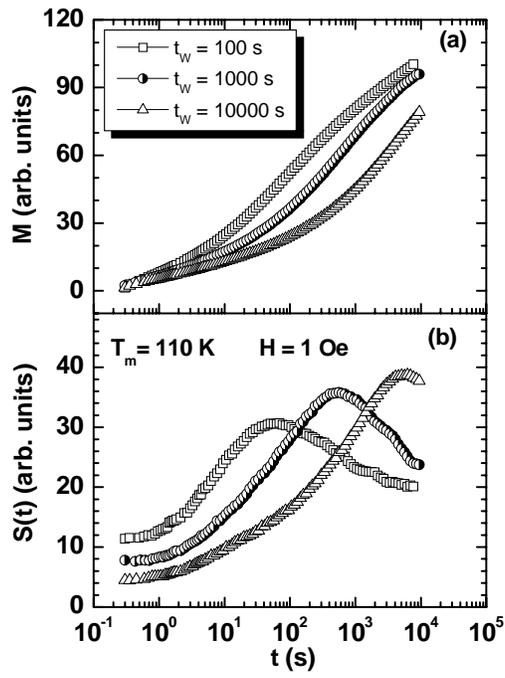

Figure 9.

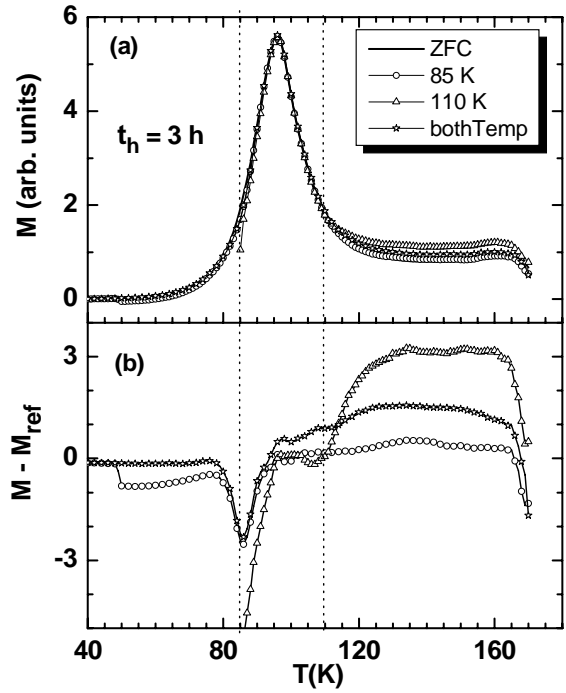

Figure 10.

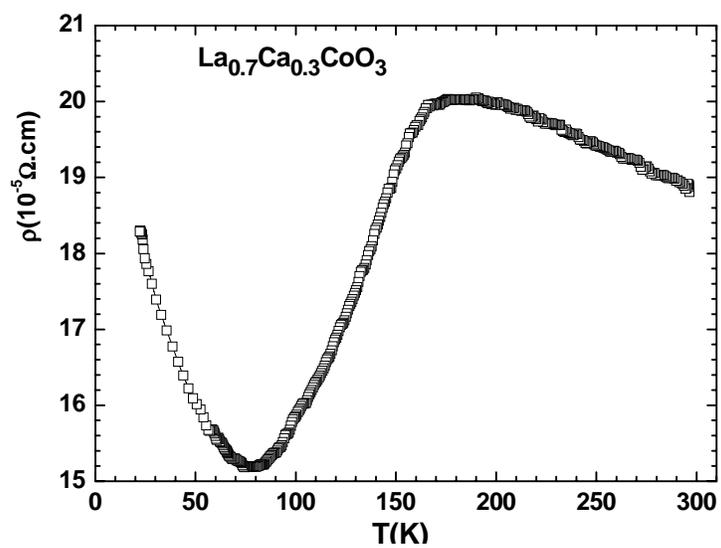

Figure 11.